\documentclass[aps,pre,floatfix,twocolumn,showpacs]{revtex4}
\usepackage{latexsym}
\usepackage{graphicx}
\newcommand{\figwidth}{3.0 in}

\usepackage{subfigure}  
\usepackage{epsfig}
\usepackage{amsmath}

\usepackage{color}
\begin{document}

\title{Flat histogram quantum Monte 
Carlo for analytic continuation to real time}
\author{Nikolaos G. Diamantis$^{1}$}
\author{Efstratios Manousakis$^{(1,2)}$}
\affiliation{
$^{(1)}$Department   of    Physics,   University    of   Athens,
  Panepistimioupolis, Zografos, 157 84 Athens, Greece \\
$^{(2)}$ Department  of  Physics and National High Magnetic Field Laboratory,
  Florida  State  University,  Tallahassee,  FL  32306-4350,  USA}

\date{\today}

\begin{abstract}
The Quantum Monte Carlo (QMC) method can yield the imaginary-time dependence
of a correlation function 
$C(\tau)$ of an
operator $\hat O$. 
The analytic continuation to real-time proceeds by 
means of a ``numerical inversion'' of these data to find the response function
or spectral 
density $A(\omega)$ corresponding to $\hat O$. Such
a technique is very sensitive to the statistical errors in $C(\tau)$ especially
for large values of $\tau$, when we are interested in the 
low-energy excitations.
In this paper, we find that if we use the flat histogram technique
in the QMC method, in such a way to make the {\it histogram of} 
$C(\tau)$ {\it flat}, the
results of the analytic continuation for low-energy excitations
improve using the
same amount of computational time. To demonstrate the idea we
select an exactly soluble version of the single-hole motion in 
the $t-J$ model and the diagrammatic Monte Carlo technique.
\end{abstract}
\maketitle

\section{Introduction}
Quantum Monte Carlo simulation, an undeniably useful tool in
addressing a number of issues  in quantum many-body physics, cannot be
used to simulate real-time dynamics. By means of analytic continuation
to Euclidean time $\tau$ ($t \to -i \tau$), however, the Schr\"odinger
equation turns into a diffusion equation, which can be simulated using
random  walks  which  explore  the  potential  landscape  by  spending
proportionately  more ``time'' near  the valleys  of the  potential and
less  ``time'' near  the potential  heights.  The  same transformation
into  imaginary time  turns the  path integral  representation  of the
evolution  operator from  an  integral over  paths of  computationally
``nasty''  phase  factors,  a   problem  almost  impossible  to  treat
stochastically, into an integral over paths in imaginary time weighted
by  a real  and  positive ``Boltzmann-like''  weight;  this weight  is
interpreted  as a  well-behaved probability  for a  particular  path to
contribute to the sum  and this interpretation allows a straightforward
stochastic treatment\cite{Ceperley}.

This  transformation,  by itself,  is  useful  because  it yields  the
interacting  ground  state  and  physical quantities  related  to  the
equilibrium statistical mechanical  description of a quantum many-body
system.  However, if we  are interested in obtaining information about
the real-time  dynamics and information  about the excitations  of the
system, an  ``inversion'' of  this ill-defined transformation  for the
results   of   correlation   functions $C(\tau)$ of  an  operator  $\hat  O$
representing a physical observable, namely,   
\begin{equation}
C(\tau)   =   \langle  
\hat T_{\tau}  \hat
O(\tau)  \hat  O^{\dagger}(0)  \rangle,
\end{equation}
 (where $\hat T_{\tau}$ is the
imaginary-time ordering operator)  from the imaginary time $\tau$ back
to real time $t$ is required.

The analytic continuation to real-time proceeds by 
means of a ``numerical inversion'' of the QMC data
on $C(\tau)$ to find the spectral 
function $A(\omega)$. These inversion techniques, such as the so-called 
maximum entropy method\cite{MEM} or its generalization, the so-called 
stochastic analytical 
inference (SAI) method\cite{SAI,Beach},
require very accurate QMC data on $C(\tau)$. Since collective phenomena
emerge at energy scales significantly smaller than the typical short-range
interaction energy scale, we are mainly interested in sampling the long 
imaginary part of such response or correlation functions. Such response 
functions obtained by QMC are noisy data and in a limited $\tau$ range. 
If we are interested in extracting the low energy excitations, this 
information hides more clearly in the long-imaginary-time evolution
of the correlation functions which is typically obscured by statistical errors.

Flat histogram methods have been very useful in classical 
systems\cite{Berg,Wang-Landau1,Wang-Landau2,Oliviera} to 
overcome problems in simulations of first order phase 
transitions, systems with rough energy landscapes, etc.
The flat histogram idea has been extended in quantum many-body systems
and, in particular, in stochastic series expansion\cite{Troyer} to overcome
the tunneling problem in first order phase transitions, 
in the continuous-time quantum Monte Carlo approach
to the impurity solver problem used in dynamical mean-field-theory 
DMFT\cite{Gull} and the diagrammatic
Monte Carlo method\cite{Diamantis13}.

The main idea presented in this paper, in simple terms, is  the following. 
We show that the flat-histogram method can be applied to the
QMC method itself to make the histogram of $C(\tau)$ 
flat for all $\tau$ by sampling the variable $\tau$, 
and keeping track of the factors in each imaginary-time interval needed to
achieve this result. In this way, we are able to compute $C(\tau)$ in 
a greater $\tau$ range with significantly smaller stochastic error. 
This approach allows us to achieve greater degree of accuracy when 
inverting the information contained in
$C(\tau)$ to find its corresponding spectral function $A(\omega)$ in the 
entire range of values of $\omega$ of our interest.

In order to demonstrate the idea in the present paper, we need to make
specific choices of i) a non-trivial quantum many-body problem, 
ii) a specific QMC method, iii) a specific correlation function $C(\tau)$, 
and iv) a method to carry out the analytic continuation. 
Using the same methods and techniques  we will calculate the spectral
function $A(\omega)$ using  QMC with and without
the application of the flat histogram method during the QMC 
runs which produce the data on $C(\tau)$. We will show that
the flat histogram QMC method is superior to a simple QMC approach in which no flat 
histogram ideas have been implemented for such
important observables. Using our past specific experience with 
models and QMC techniques, we choose  the problem of the Green's function 
$G_{\vec k}(\tau)$ of a
 single hole in the $t-J$ model\cite{Liu1,Liu2} with the diagrammatic Monte
Carlo (DMC) method\cite{Polaron-DMC1,Polaron-DMC2,Diamantis13} using the method of the stochastic analytical
inference\cite{SAI} for the analytic continuation to 
obtain the spectral function 
$A_{\vec k}(\omega)$ from the QMC data obtained for $G_{\vec k}(\tau)$. 

We apply the flat histogram idea  with the combination of the DMC
method and the Wang-Landau method, which we will refer to as the 
flat histogram diagrammatic Monte Carlo (FHDMC) method\cite{Diamantis13}.
We will also use the
standard implementation of the DMC\cite{Polaron-DMC1,Polaron-DMC2} where a guidance function
is used when sampling $G_{\vec k}(\tau)$. In the latter case, as we will see, 
the use of a parameter $\mu$ effectively makes the histogram of 
$G_{\vec k}(\tau)$ flat. On the other hand, we will also carry out DMC simulations
using  $\mu=0$ without the application of  the flat histogram idea, 
which we will refer to as DMC0 which yields a histogram of 
$G_{\vec k}(\tau)$  very far from being flat, 
and we compare these results with those obtained with  
above two QMC methods, i.e., with standard-DMC and FHDMC. 
This comparison is made in order to show the main point of our 
paper that if the flat histogram
idea is incorporated in any QMC method,  it will provide a more accurate 
analytic continuation to real time. 

The paper is organized as follows. In Sec.~\ref{method} we describe
the problem, the model, and the general approach which we will follow.
Sec.~\ref{details} describes the computational details and our implementation 
of the analytic continuation technique which we adopted.
In Sec.~\ref{results} we present the results of the spectral function 
obtained with the DMC0, the standard DMC, and FHDMC method, 
which are compared to what we believe to be the 
``exact'' results of a soluble but restricted version of the $t-J$ model.
Lastly, in Sec.~\ref{conclusion}
we present the main conclusions of the paper.

\section{The method}
\label{method}
In order to be precise we take the example where the
operator $\hat O^{\dagger}$ is the single particle 
creation operator $a^{\dagger}_{\vec k}$, 
in which case $C(\tau)$ becomes the single particle Green's 
function $G_{\vec k}(\tau)$ in imaginary time. 
In this case the spectral function $A_{\vec k}(\omega)$ is related to 
$G$ as follows
\begin{eqnarray}
G_{\vec k}(\tau) = \int d\omega K(\tau,\omega) A_{\vec k}(\omega),
\label{eq.1}
\end{eqnarray}
where the so-called kernel $K(\tau,\omega)$ is simply $e^{-\omega\tau}$.
The spectral function is a non-negative quantity normalized to unity.

We consider a finite imaginary time range, $0<\tau < \tau_{max}$ and we divide
it into $L$ equal intervals. By integrating Eq.~\ref{eq.1}
in each time interval $i$ we obtain
\begin{eqnarray}
{\cal G}_{\vec k}(i) &=& \int d\omega {\bar K}(i,\omega) A_{\vec k}(\omega), 
\label{eq.2a}\\
{\cal G}_{\vec k}(i) &\equiv& {1 \over {\Delta\tau_i}}  \int_{\tau_{i-1}}^{\tau_i}
G_{\vec k}(\tau) d\tau, \label{eq.2b}
\end{eqnarray} 
and ${\bar K}(i,\omega)$ is the average value of the 
kernel 
$K(\tau,\omega) = e^{-\omega \tau}$ in the $i$ interval, i.e.,
${\bar K}(i,\omega) = {{e^{-\omega \tau_i}} \over {\omega \Delta \tau_i}}
[e^{\omega \Delta \tau_i} - 1 ]$.

In Ref.~\onlinecite{Diamantis13}, we have shown that we can apply the 
flat histogram technique on the so-called diagrammatic
Monte Carlo method\cite{Polaron-DMC1,Polaron-DMC2,Mishchenko1,Mishchenko2,Mishchenko3,Mishchenko4} 
to make the histogram ${\cal G}_{\vec k}(i)$ of
$G_{\vec k}(\tau)$  flat. The idea was demonstrated
on the Fr\"ohlich polaron problem. The results of the flat histogram
DMC (FHDMC) on the estimate for the polaron ground state were significantly 
better than the result of the DMC0. However, as argued in the
paper\cite{Diamantis13} the 
polaron problem spectrum was characterized by a gap, and the full 
advantage of the flat histogram method over DMC0 
could not fully demonstrated.


In the present paper we use a simplified version of the $t-J$ model, in which 
the Heisenberg and the hole-hopping terms are linearized within the
spin-wave approximation to obtain a polaron-like Hamiltonian\cite{KLR,Liu2}, 
i.e.,    
\begin{eqnarray}
\hat H &=& - \sum_{\vec k,\vec q} 
g(\vec k, \vec q) a^{\dagger}_{\vec k+\vec q} a_{\vec k} b_{\vec q} + H.c. \nonumber \\
&+& \sum_{\vec k} \hbar \omega(k) b^{\dagger}_{\vec k} b_{\vec k}, \\
g(\vec k, \vec q) &=& {{4 t } \over {\sqrt{N}}} (u_{\vec q} \gamma_{\vec k- \vec q} + v_{\vec q} \gamma_{\vec k}), \\
\gamma_{\vec k} &=& - 2 t (\cos(k_x a) + \cos(k_y a)),
\label{eq.3} 
\end{eqnarray}
where the operator $b^{\dagger}_{\vec q}$ is the Bogoliubov spin-wave 
creation operator, $\omega(k)$ is the spin-wave dispersion
of the square lattice quantum antiferromagnet\cite{RMP} and 
$a^{\dagger}_{\vec k}$ is
the hole creation operator. Here $g(\vec k,\vec q)$ is the coupling of the 
hole to spin waves and $u_{\vec k}$ and $v_{\vec k}$ are the coefficients of the
the  Bogoliubov transformation as given in Ref.~\onlinecite{RMP}

The single-hole spectral function of the above Hamiltonian
can be approximated by the non-crossing approximation(NCA)\cite{KLR,Liu1,Liu2}.
The diagrammatic Monte Carlo (DMC) method with or 
without\cite{Mishchenko1,Mishchenko2,Mishchenko3,Mishchenko4} the
incorporation of the flat histogram technique can be applied to 
the problem of a single-hole. 
In this paper in order to demonstrate the method, we wish to restrict our DMC
to the sampling of only the diagrams contributing to the NCA. This 
restriction is
actually a more difficult approach to numerically implement than the one 
in which the Markov process samples all the connected diagrams contributing
to $G_{\vec k}(\tau)$.  
The reason
for restricting ourselves within the NCA diagrammatic space,
is because in this case there is an ``exact'' solution to the 
problem which we can use to measure the success of the
approach discussed here. 
In order to obtain the ``exact'' solution, we have 
recalculated the single-hole Green's function
within the NCA approach described in  Ref.~\onlinecite{Liu2}
for $\vec k = (\pi/2,\pi/2)$  for 
a $32\times 32$ size square lattice and a finite
imaginary part was used in the free-hole propagator with $\epsilon=0.002$
(in units of the hopping parameter $t$)
which smears the lowest energy quasiparticle peak.
Note that as discussed 
in Ref.~\onlinecite{Liu2} on this size lattice the finite-size
effects were found to be small. 
The results of these
calculations will be used as default models and we will
refer to them as ``exact'' solutions.

We have applied the DMC0, the standard DMC, and FHDMC methods to obtain 
$G_{\vec k}(\tau)$ for $\vec k = (\pi/2,\pi/2)$ and $J/t=0.2$ in the  NCA
restricted diagrammatic space.
 We have
carried out QMC simulations for a range of $J/t$ to compare
with the exact results in order to make sure that our computer
programs are correct.  

Now imagine that we obtain a set of data $G^{(d)}_{\vec k}(i,j)$ 
$i=1,...,L$, $j=1,...,N_d$ on
${\cal G}_{\vec k}(i)$ obtained by any of the previously discussed 
three different QMC methods, where the $j$ denotes the data bin.
For the analytic continuation we will use a generalization of the maximum 
entropy method, the so-called
stochastic analytical inference (SAI) technique. In the later approach,  
$A_{\vec k}(\omega)$ is obtained as the  average over all its  possible
forms in a Monte Carlo integration where the particular form of
$A_{\vec k}(\omega)$ is selected from a distribution determined by the
so-called {\it default model} $D_{\vec k}(\omega)$ and 
the probability of any proposed form of
$A_{\vec k}(\omega)$ to be the true form given the input data 
$G^{(d)}_{\vec k}(i,j)$ is given by 
\begin{eqnarray}
P[A/G^{(d)}_{\vec k}] &\sim& e^{-{{\chi^2[A]} \over {2 \alpha}}},
\label{eq.4}
\end{eqnarray}
where $\chi^2[A]$ is the $\chi^2$ determined from the data 
$G^{(d)}_{\vec k}(i,j)$ on ${\cal G}_{\vec k}(i)$ and the result which
is obtained from Eq.~\ref{eq.1} using the proposed form of
$A_{\vec k}(\omega)$. In the evaluation of $\chi^2[A]$ the
covariance matrix of the data is used.  
The technical details of how one {\it selects} the particular
proposed $A_{\vec k}(\omega)$ according to a given default model $D(\omega)$
and other important details of this calculation are discussed in
Sec.~\ref{details}. 

In the process of the analytic continuation we will analyze the
QMC data obtained with the three different QMC methods which we 
discussed above.
Namely, first we will use DMC0 approach in which no guidance function
has been implemented (i.e., $\mu=0$). Second, we will use data obtained by  means
of standard-DMC (SDMC) where the histogram of $G(\tau)$ is
 made approximately flat
by multiplying $G(\tau)$ with an exponential ``guidance'' function (we use the
same terminology used in Ref.~\cite{Polaron-DMC1}) of the form 
$\exp(\mu \tau)$. In Fig.~\ref{fig:1}(a) we compare the 
histograms of $G(\tau)$ obtained with both DMC0 and standard-DMC. 
It is clear that the standard DMC makes the histogram of $G(\tau)$ approximately
flat. Therefore, we will consider
this method as an approximately flat histogram approach.
Last, we will also use data obtained by the FHDMC. Notice that
the histogram of $G(\tau)$ which we obtain using the Wang-Landau algorithm
is flat to better than $95\%$. 

Furthermore, we will use three different default models  in each of
the above cases, i.e., the total number of results presented will correspond 
to nine combinations of QMC data and default models. In one series of results
presented in this paper, we will use as default model
the ``exact'' solution to the problem. As a second choice for default
model we will use a flat distribution, and as  a third choice we will use a
default model which shares some aspects of the ``exact'' solution
but it differs significantly from the ``exact'' solution
and we call it the ``incorrect'' default model.
These three choices are discussed in Sec.~\ref{defaults} in detail.

\section{Computational Details}
\label{details}

As discussed in the previous Section, we will restrict our DMC approach
to sample the diagrammatic subspace spanned only by the diagrams
included in the non-crossing approximation\cite{KLR,Liu1,Liu2}.
The only reason for this restriction is that the problem of 
the single-hole spectral function can be solved ``exactly'' (``exact'' 
with the limitations discussed in the previous Section) and, this, 
can be used to test the accuracy of the results of QMC method.

\begin{figure}[htp]
\vskip 0.4 in
     \begin{center}
        \subfigure[]{%
            \label{fig:1a}
            \includegraphics[width=\figwidth]{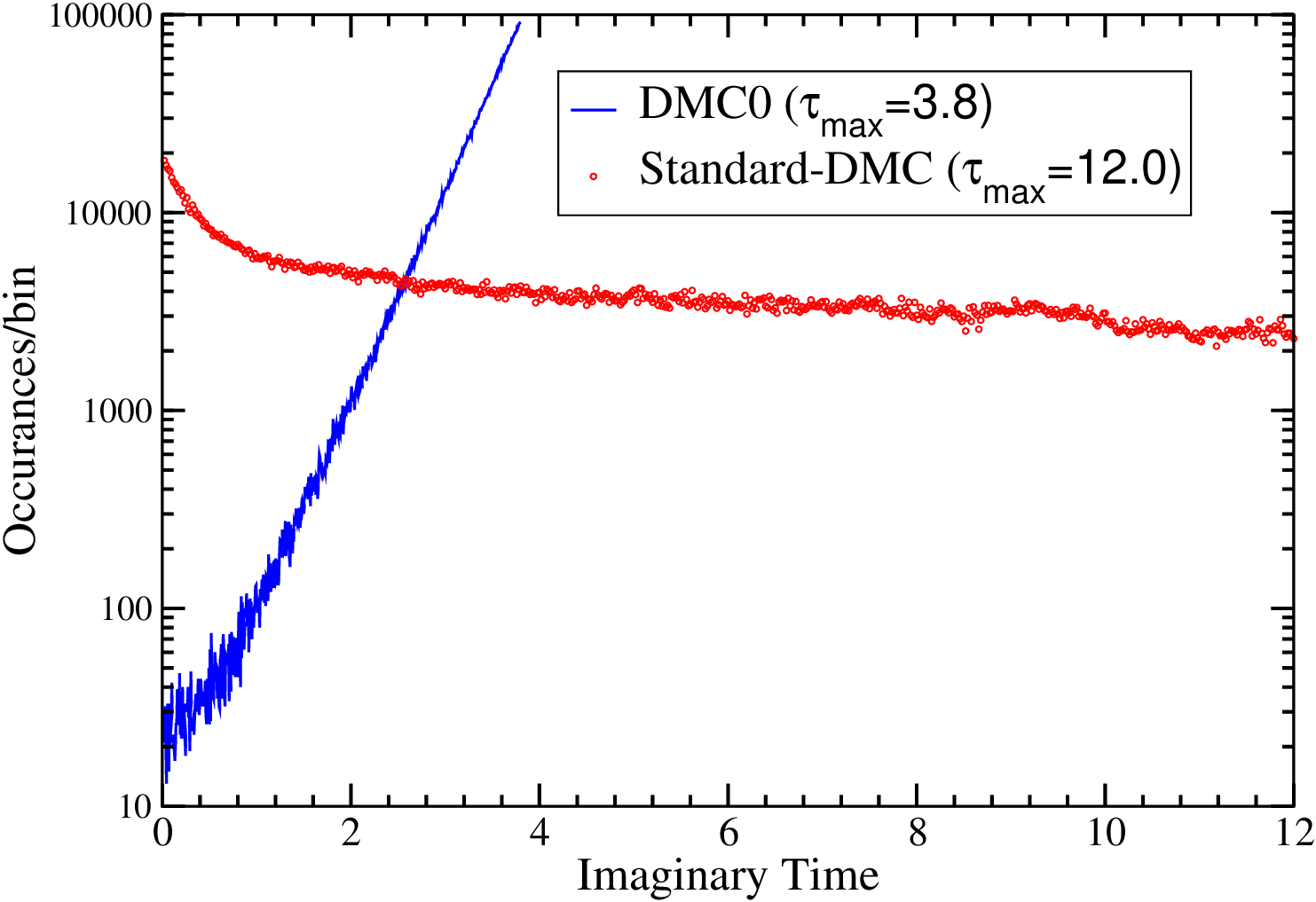}
        }%
\vskip 0.2 in
        \subfigure[]{%
            \label{fig:1b}
            \includegraphics[width=\figwidth]{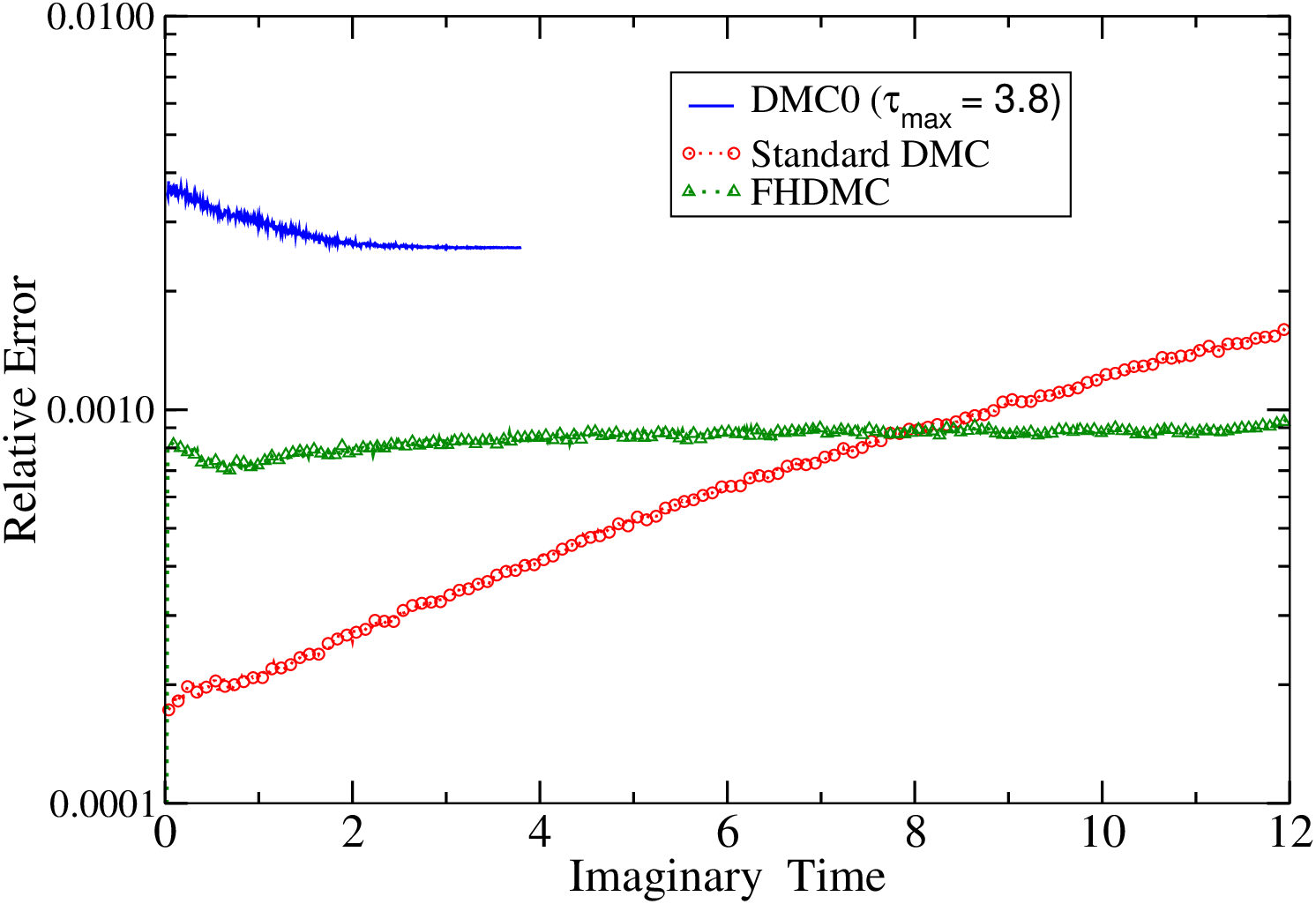}
        }%
    \end{center}
\caption{(Color on-line) (a) Our results for the histogram of 
 $G_{\vec k}(\tau)$ (with $\vec k=(\pi/2,\pi/2)$) as a function
of the imaginary time as obtained with  the standard-DMC
and the DMC0. In the case of FHDMC the histogram is made through the
Wang-Landau algorithm flat to a high degree of precision. 
(b) Our results for the relative error 
as a function
of the imaginary time $\tau$ as obtained with DMC0, standard-DMC, and FHDMC
for the approximately the same amounts of CPU time.}
\label{fig:1} 
\vskip 0.2 in
\end{figure}

Furthermore,  we will work with three sets of QMC data obtained for the
histogram of $G_{\vec k}(\tau)$ for $\vec k = (\pi/2,\pi/2)$ and
 $J/t=0.2$.  
One set is obtained with DMC0, the second set is obtained with standard DMC, 
and the third set is obtained with FHDMC; all three sets of data are obtained with
 approximately the same amount of CPU time in order to compare. 
The first set of data for the histogram of $G_{\vec k}(\tau)$ is obtained for
$0 < \tau < 3.8$ (in units of the inverse hopping matrix element $t$)
while the other two sets of data are obtained for
$0 < \tau < 12$ and in all three sets of data we have used 600 $\tau$-intervals ($L=600$) and the number of
data bins $N_d=1200$ for each time-interval. 
In Fig.~\ref{fig:1}(a) we show that the histogram of $G(\tau)$ obtained
by the standard DMC is fairly flat over a very wide range of $\tau$
while that of the DMC0 is approximately an exponentially increasing 
function of $\tau$. Therefore, we will consider that the
standard DMC method is a method which produces an approximately flat
histogram.

 The relative statistical error
of the FHDMC results is more-or-less independent of $\tau$ (see also 
Fig.~\ref{fig:1}(b)) and it was approximately $\sigma_r = 8 \times 10^{-4}$
while for the results obtained using standard-DMC, $\sigma_r$ ranges
from $2\times 10^{-4}$ for small $\tau$ up to $2 \times 10^{-3}$ for large
$\tau$. The relative error for the DMC0 results 
is significantly larger and of the
order of $3 \times 10^{-3}$ (see Fig.~\ref{fig:1}(b)).
These relative errors were obtained by using approximately
the same amount of CPU time such that the comparison between the three
DMC methods to be meaningful.

\subsection{Application of the Stochastic Analytical Inference method}

Since our data on  $G_{\vec k}(\tau)$ obtained by either DMC0, standard DMC or FHDMC 
are for a specific fixed value of $\vec k$, they will be simply denoted
as $ G^{(d)}(i,j)$ and the histogram of $G_{\vec k}(\tau)$  will be
denoted as ${\cal G}(i)$,
and $A(\omega)$ is an abbreviation for $A_{\vec k}(\omega)$ for 
fixed $\vec k$.
There are $N_d$ data bins $G^{(d)}(i,j)$, $j=1,2,...,N_d$ obtained for $G(i)$
in each imaginary time slice $i$ and $i=1,2,...,L$.

Because as we produce the data table $G^{(d)}(i,j)$ the imaginary time
$\tau$ is sampled during any of the three DMC simulations, there is
correlation between the data for the different imaginary time intervals.
Therefore, in order to compute $\chi^2$ we need the data
covariance matrix ${\bf C}$ defined from its matrix elements as follows:
\begin{eqnarray}
C(k,l) &=& {1 \over {N_d(N_d-1)}} \sum_{j=1}^{N_d}
[\bar {G}^{(d)}(k)-G^{(d)}(k,j)] \nonumber \\
&\times& [\bar {G}^{(d)}(l)- G^{(d)}(l,j)], \hskip 0.1 in k,l=1,2,...,L,
\end{eqnarray}
where $\bar {G}^{(d)}(k)$ is the average of the data for the $k$ time-slice,
i.e, 
\begin{eqnarray}
\bar {G}^{(d)}(k) = {1 \over {N_d}} \sum_{j=1}^{N_d}
G^{(d)}(k,j).
\end{eqnarray}
In terms of ${\bf C}$, $\bar {G}^{(d)}$ and the proposed $A(\omega)$ which
gives a ${\cal G}(i)$ via Eq.~\ref{eq.1}, the $\chi^2$ is given as
\begin{eqnarray}
\chi^2 = (\bar {{\bf G}}^{(d)}-{\bf G})^T {\bf C}^{-1} (\bar {{\bf G}}^{(d)}-
{\bf G}),
\label{chis1}
\end{eqnarray}
where $(\bar {{\bf G}}^{(d)})^T \equiv (\bar {G}^{(d)}(1),\bar {G}^{(d)}(2),...,
\bar {G}^{(d)}(L))$ and ${\bf G}^T\equiv({\cal G}(1),{\cal G}(2),...,{\cal G}(L))$.   

We determine the orthogonal matrix ${\bf O}$ and the diagonal matrix 
${\bf d}$ such
that:
\begin{equation}
{\bf C} = {\bf O} {\bf d} {\bf O}^T.
\end{equation}
Then, we can simply write $\chi^2$ as follows:
\begin{eqnarray}
\chi^2 &=& ({\bar {\bf G}}^{(d)}_f-{\bf G}_f)^T({\bar {\bf G}}^{(d)}_f-
{\bf G}_f), \label{chis2}
\\
{\bar {\bf G}}^{(d)}_f &=& {\bf d}^{-1/2} {\bf O}^T {\bar {\bf G}}^{(d)}, \\
 {\bf G}_f &=& {\bf d}^{-1/2} {\bf O}^T {\bf G}.
\end{eqnarray}

In the SAI method
$A(\omega)$ is obtained as the  average over all its  possible
forms by a Metropolis Monte Carlo sampling, where the particular form of
$A(\omega)$ is selected from a distribution determined by the
so-called {\it default model} $D(\omega)$ and 
the probability of any proposed form of
$A(\omega)$ to be the true one, given the input data 
$G^{(d)}(i,j)$, is given by Eq.~\ref{eq.4}.
The $\chi^2[A]$ is obtained using Eq.~\ref{chis2} 
determined from the data 
$G^{(d)}(i,j)$ on  ${\cal G}(i)$ and ${\cal G}(i)$ which are determined from the 
assumed values of
$A(\omega)$.  The optimum choice of the ``temperature'' $\alpha$
is made according to the discussion in 
Refs.~\onlinecite{Sandvik,Beach,Syljuasen}.

We apply the Metropolis algorithm using the expression given by 
Eq.~\ref{eq.4} as the acceptance probability and we calculate
the average spectral function $A(\omega)$. In the application of the
Metropolis algorithm we use as selection probability of a particular
$A(\omega)$, a distribution which is related to the so-called
default model $D(\omega)$, which 
contains our prior knowledge about the actual $A(\omega)$. 
We consider the frequency interval $[\omega_{min},\omega_{max}]$
and we assume that $A(\omega)$ is zero outside this interval.
First, we slice this frequency interval into $N_w$ intervals
$\Delta\omega(i)$ around a middle-frequency $\omega(i)$, $i=1,2,...,N_w$.
We define a normalized histogram based on the default model $D(\omega)$ as
\begin{eqnarray}
{\cal D}(i)&=& {{D(\omega(i))} \over {\sum_{j=1}^{N_w} D(\omega(j))
\Delta\omega(j)}}, 
\end{eqnarray}
and based on it we define a new variable $x(i)$ which takes
values in the interval $[0,1]$, which is sliced in intervals
$[x(i-1),x(i)]$ of width 
\begin{eqnarray}
\Delta x(i) = {\cal D}(i) \Delta \omega(i), \hskip 0.3 in i=1,2,...,N_w,
\end{eqnarray}
such that we can define the following
\begin{eqnarray}
x(0) = 0, \hskip 0.2 in x(i) = \sum_{j=1}^i \Delta x(j), 
\hskip 0.2 in x(N_w)=1.
\end{eqnarray}
We select normalized ``configurations'' of $n(x)$ ($n(x)\ge 0$, 
$\int_0^1 n(x) dx =1$) from the uniform 
distribution\cite{Beach,SAI}. The height ${\cal A}(i)$ of the
histogram of $A(\omega)$ in the interval  $\Delta\omega(i)$ of $\omega$
is obtained as
\begin{eqnarray}
{\cal A}(i) = { 1 \over {\Delta\omega(i)}} \int_{x(i-1)}^{x(i)} n(x) dx.
\end{eqnarray}


\subsection{Choice of the default models}
\label{defaults}

In order to invert these data using the SAI method discussed
in the previous subsection, we need a default model. We are going 
to use the following three default models.

(a) First, we will use the ``exact'' default
model, i.e., the solution obtained within NCA by solving the
Dyson's equation self-consistently as was done in Ref.~\cite{Liu1,Liu2}.
For this purpose we have re-calculated the spectral function
for $\vec k = (\pi/2,\pi/2)$ for $J/t=0.2$ on a $32\times 32$ size
lattice and $\epsilon/t =0.002$. This result is close to the
``exact'' NCA solution, but still has some small finite-size effects
and finite-$\epsilon$ effects.

(b) A a second choice, we will use as a default model a flat distribution,
i.e., 
\begin{eqnarray}
A(\omega) = \left \{ \begin{array}{ccc} 
C, & \omega_a < \omega < \omega_b, \\
0, & \mathrm{otherwise},   \end{array} \right \}
\label{flat}
\end{eqnarray}
where $C=1/(\omega_b-\omega_a)$ and $\omega_a = -3$ and  $\omega_b = 5$.

(c)  A third default model, which we will use is the following: 
Using the  approach described in 
Ref.~\onlinecite{Liu2} we also recalculated 
the spectral function solution for $A_{\vec k}(\omega)$ for the same 
value of $J/t=0.2$ within NCA but for $k=(0,0)$. This spectral function 
will be used as an option for a default model in the SAI method.
The reason we would like to use
this default model is to investigate the extent to which this method is
capable of finding the correct solution starting from a default model which 
shares some features with the correct solution, such the location of the
peaks but not others, such as the ``strength'' of the peaks. This
could be somewhat similar with a practical case scenario, where, 
for example, let us assume that we have performed a GW 
density-functional-theory 
based calculation for the quasiparticle spectral function of a real material, 
and,  we  have obtained more-or-less correct energy eigenvalues, but the
quasiparticle wavefunctions are not close to the correct ones. 
In this case the peaks will be close to the correct positions
but the spectral weights should be different from the actual ones.
In principle, we can imagine that one starts a QMC simulation of a real 
material using as a default model
the spectral function obtained in such a GW calculation.

\section{Results}
\label{results}

\begin{figure*}[htp]
\vskip 0.2 in
     \begin{center}
        \subfigure[]{%
            \label{fig:2a}
            \includegraphics[width=\figwidth]{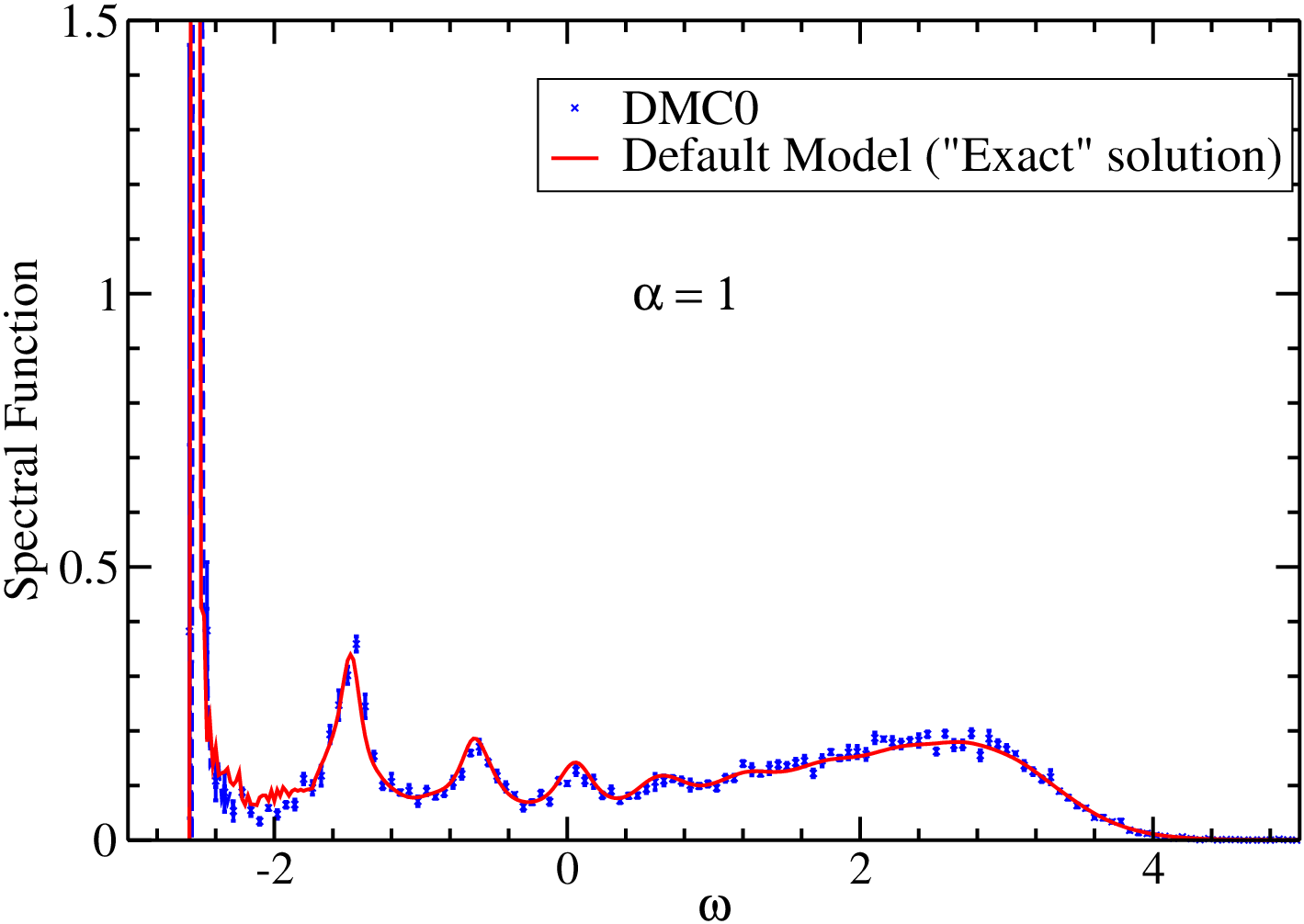}
        }%
\hskip 0.2 in        \subfigure[]{%
            \includegraphics[width=\figwidth]{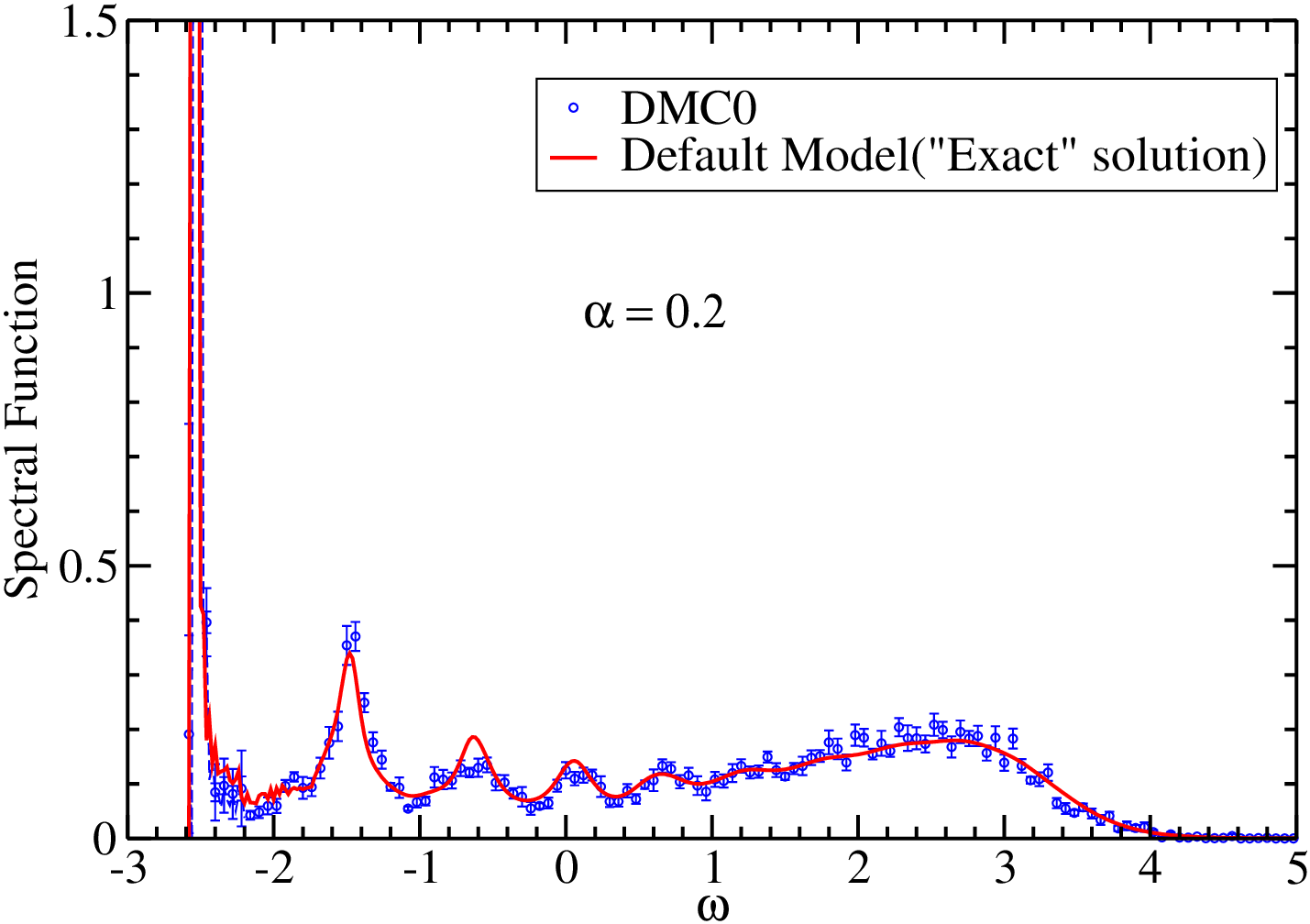}
            \label{fig:2b}
        }%
\\ \vskip 0.1 in

        \subfigure[]{%
            \label{fig:2c}
            \includegraphics[width=\figwidth]{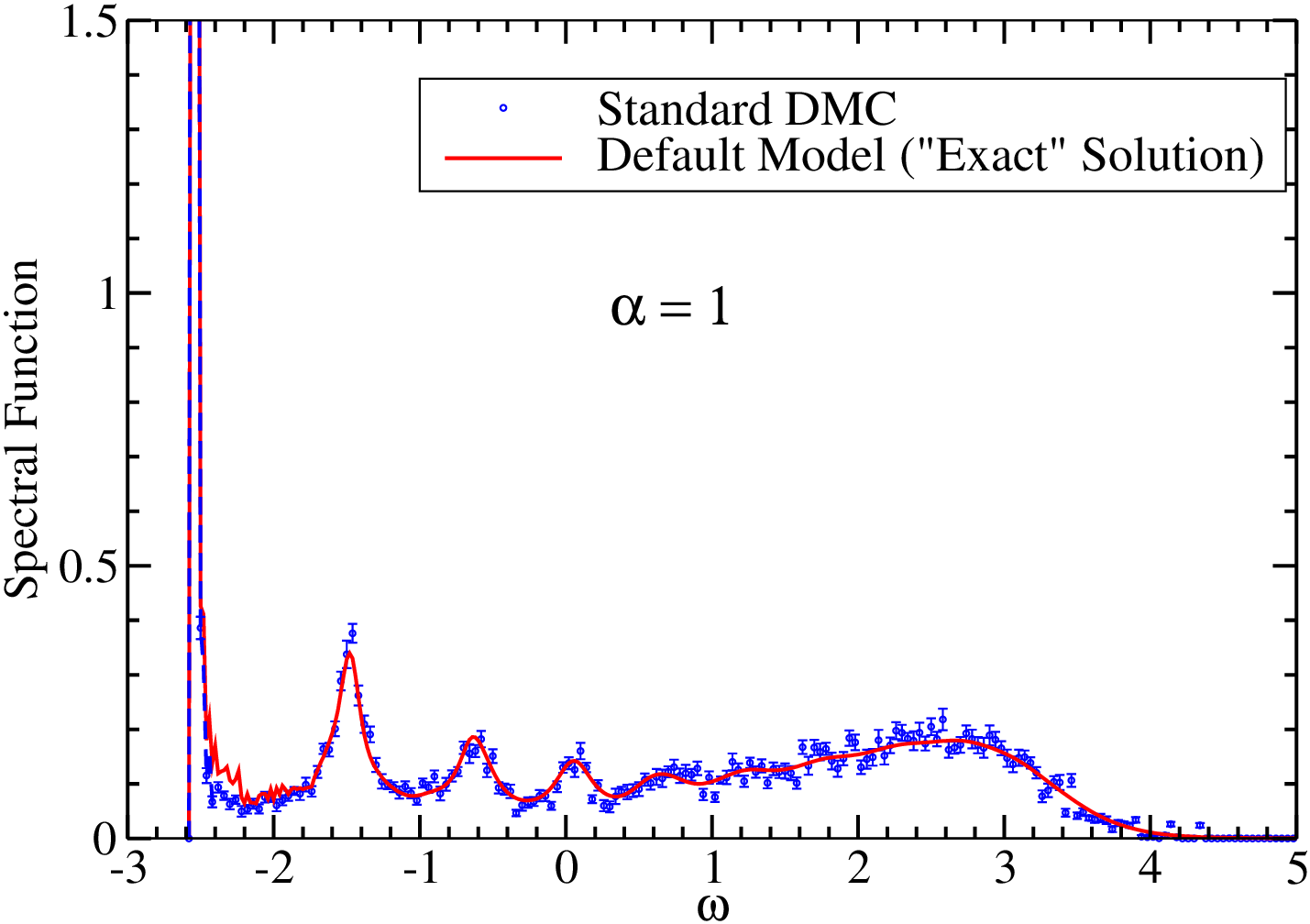}
        }%
\hskip 0.2 in        \subfigure[]{%
            \includegraphics[width=\figwidth]{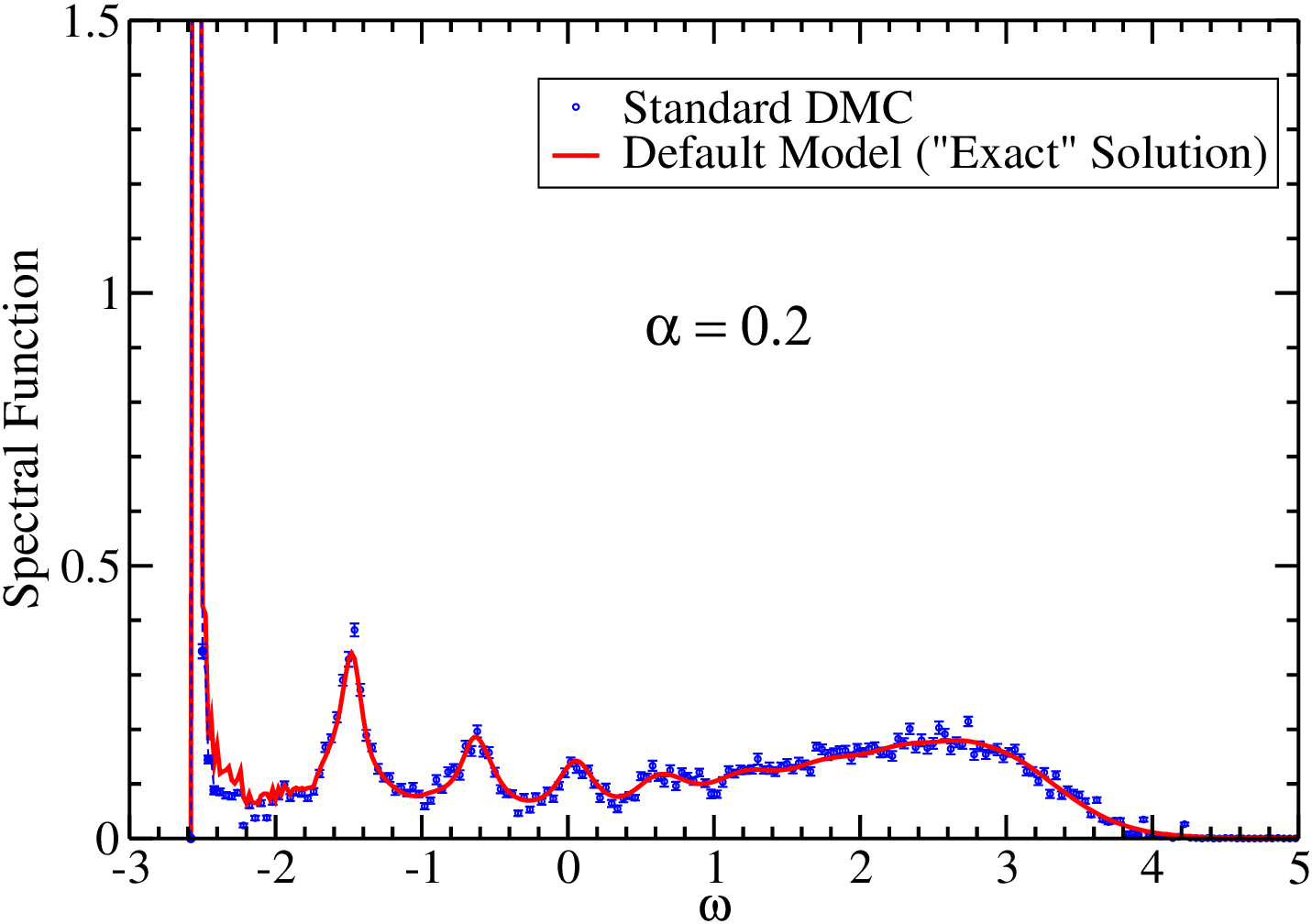}
            \label{fig:2d}
        }%
\\ \vskip 0.1 in

        \subfigure[]{%
            \label{fig:2e}
            \includegraphics[width=\figwidth]{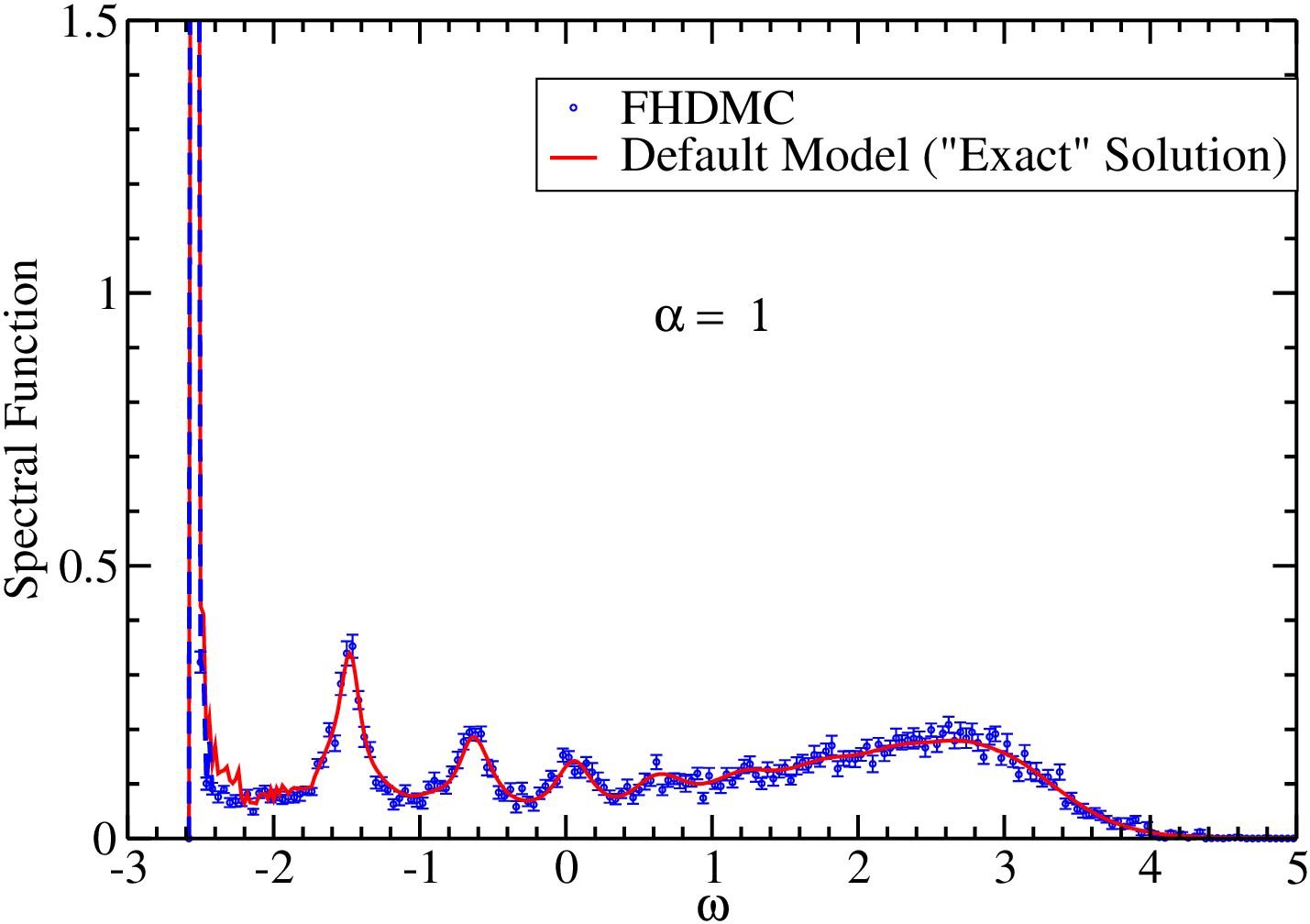}
        }%
\hskip 0.2 in        \subfigure[]{%
            \includegraphics[width=\figwidth]{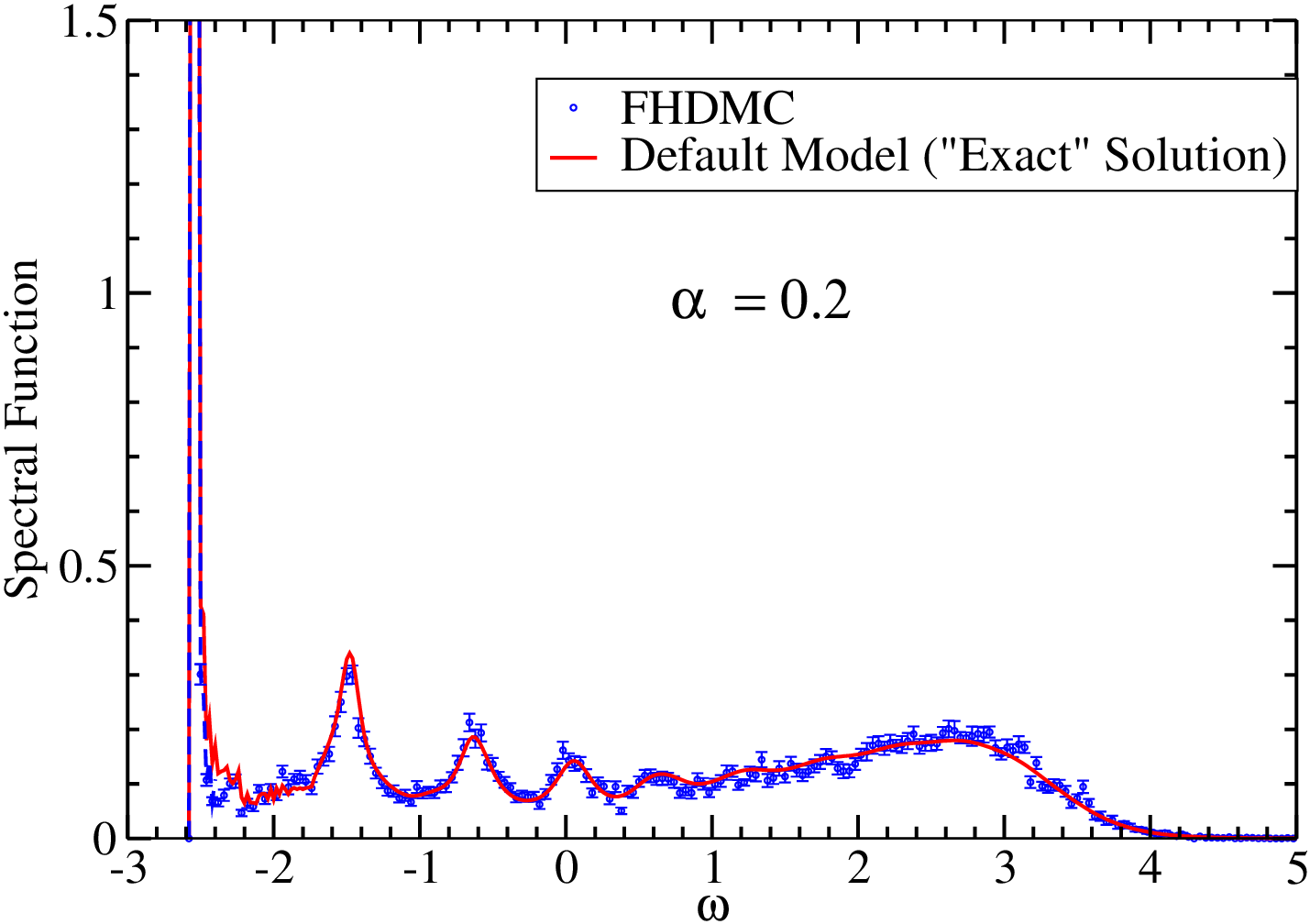}
            \label{fig:2f}
        }%

    \end{center}
\caption{(Color on-line) Analytic continuation of FHDMC and standard DMC data of the
$t-J$ model for $J/t=0.2$ for $\vec k=(\pi/2,\pi/2)$ 
obtained from the SAI technique using 
as default model the ``exact'' solution. 
Subfigures (a) and (b) are obtained using
the DMC0 data for $\alpha=1$ and $\alpha=0.2$ respectively.
Subfigures (c) and (d) are obtained using
the standard DMC data for $\alpha=1$ and $\alpha=0.2$ respectively.
Subfigures (e) and (f) are obtained using
the FHDMC data for $\alpha=1$ and $\alpha=0.2$ respectively.
}
\label{fig:2} 
\vskip 0.3 in
\end{figure*}

\begin{figure}[htp]
\vskip 0.3 in
     \begin{center}
        \subfigure[]{%
            \label{fig:3a}
            \includegraphics[width=\figwidth]{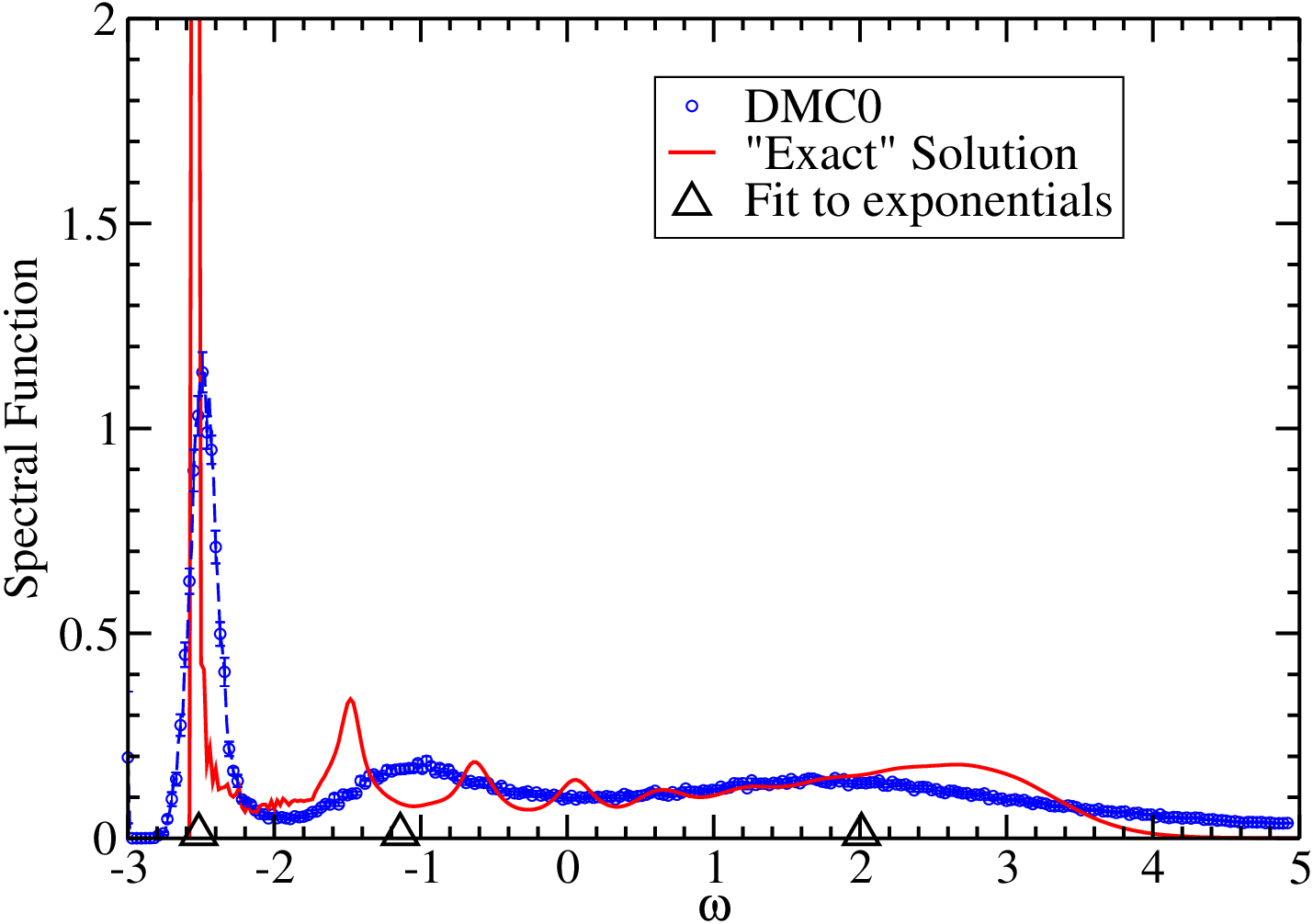}
        }%
\\ \vskip 0.1 in

        \subfigure[]{%
            \label{fig:3b}
            \includegraphics[width=\figwidth]{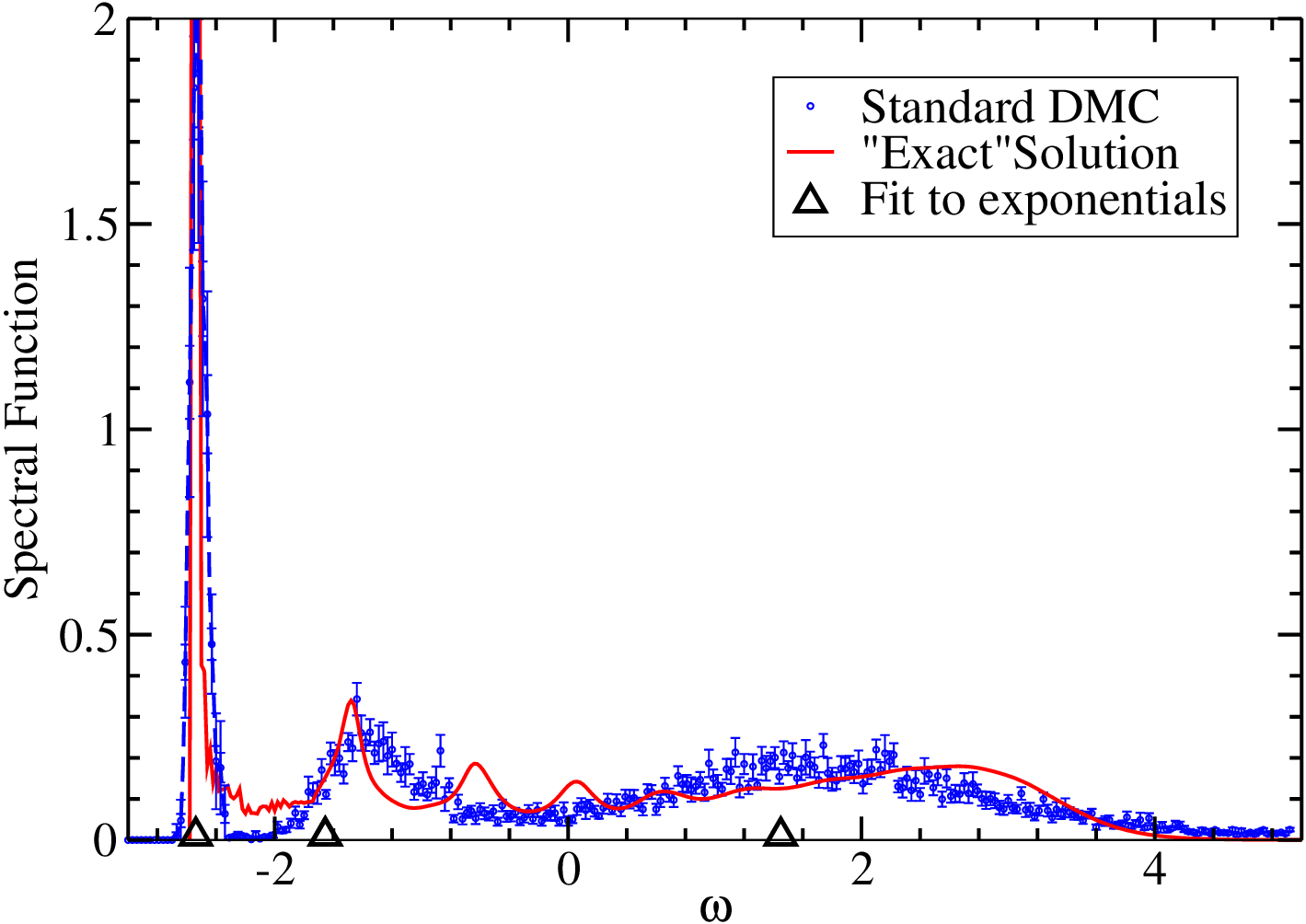}
        }%
\\ \vskip 0.1 in

        \subfigure[]{%
            \label{fig:3c}
            \includegraphics[width=\figwidth]{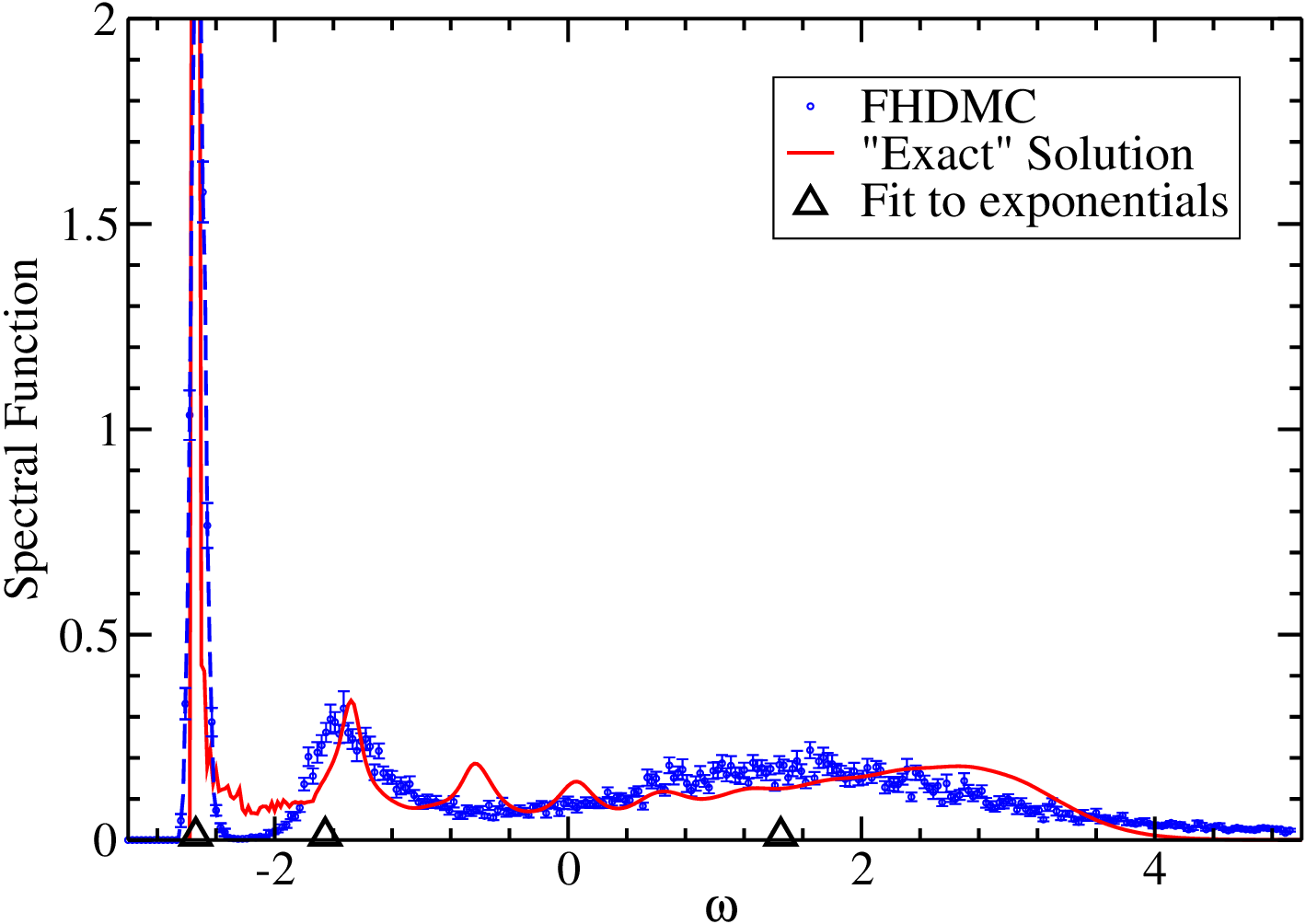}
        }%

    \end{center}
\caption{(Color on-line) Analytic continuation of QMC data 
obtained from the SAI technique using as default model a flat histogram
and $\alpha=1$.
Subfigure (a) is obtained using
the DMC0 data.
Subfigure (b) is obtained using
the standard DMC data.
Subfigure (c) is obtained using
the FHDMC data.
 The black triangles pointing up show the values of the frequencies 
obtained by a fit to the data
using three exponentials.}
\label{fig:3} 
\vskip 0.3 in
\end{figure}

\begin{figure}[htp]
\vskip 0.4 in
     \begin{center}
        \subfigure[]{%
            \label{fig:4a}
            \includegraphics[width=\figwidth]{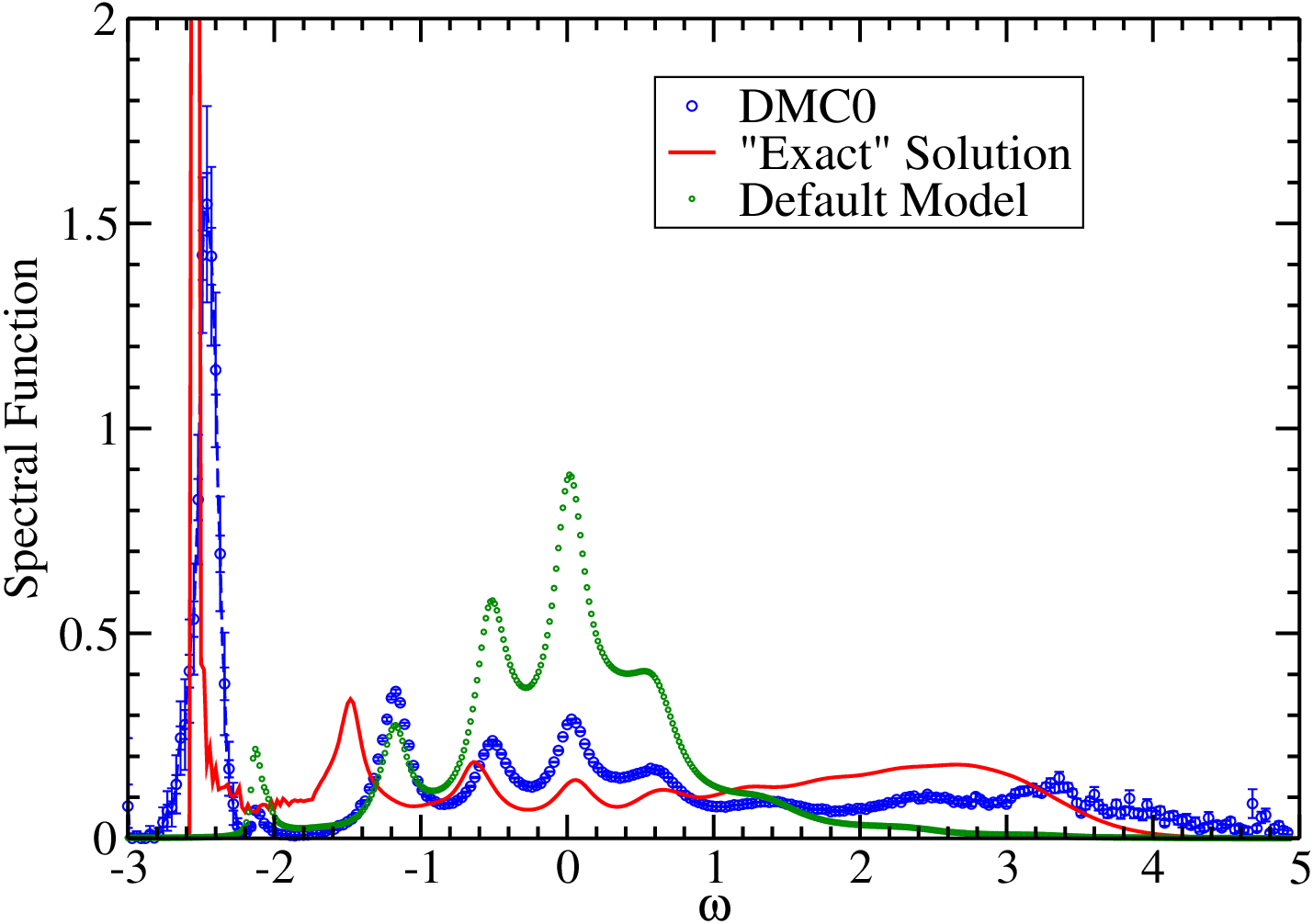}
        }%
\vskip 0.2 in        \subfigure[]{%
            \label{fig:4b}
            \includegraphics[width=\figwidth]{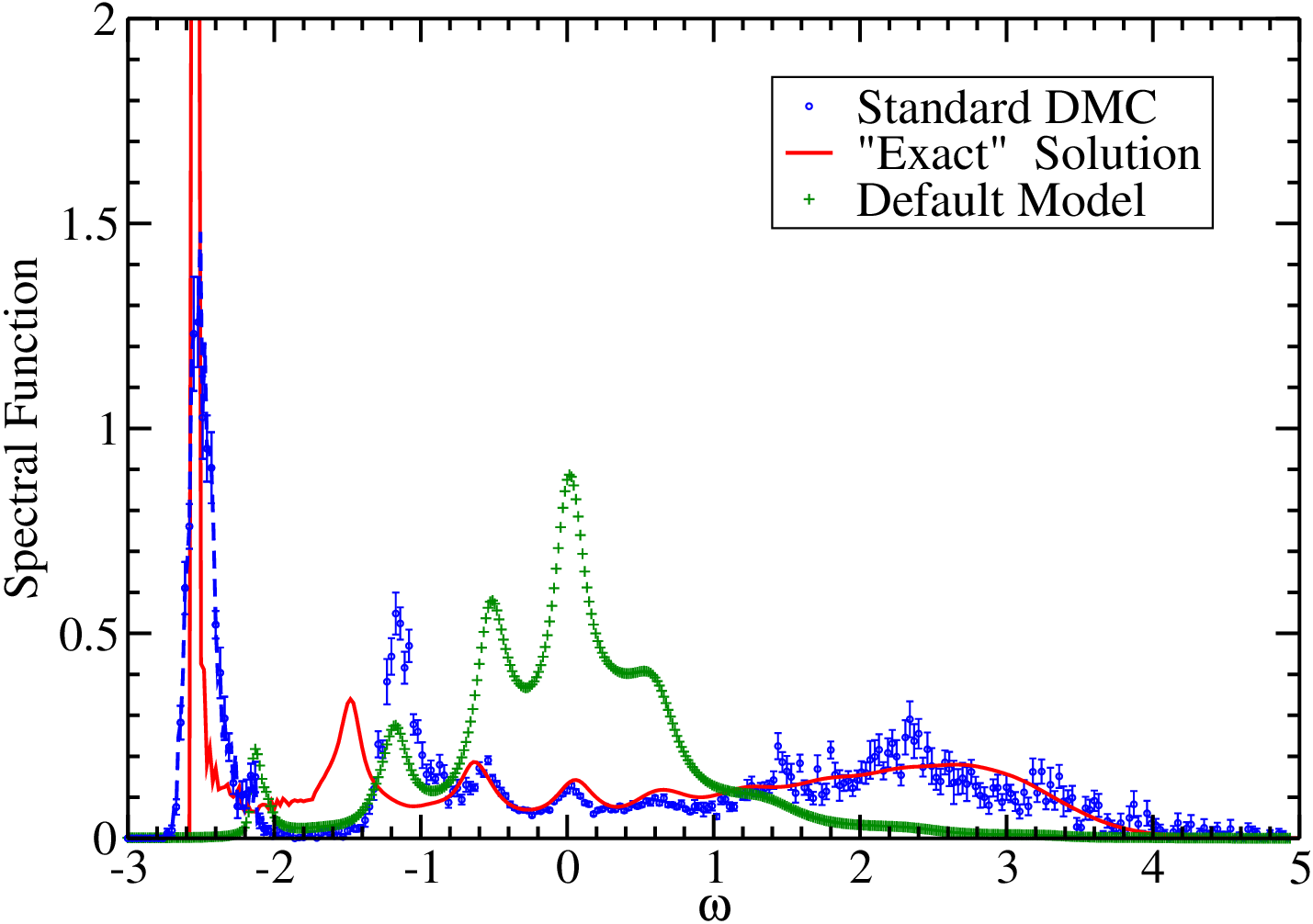}
        }%
\vskip 0.2 in        \subfigure[]{%
            \label{fig:4c}
            \includegraphics[width=\figwidth]{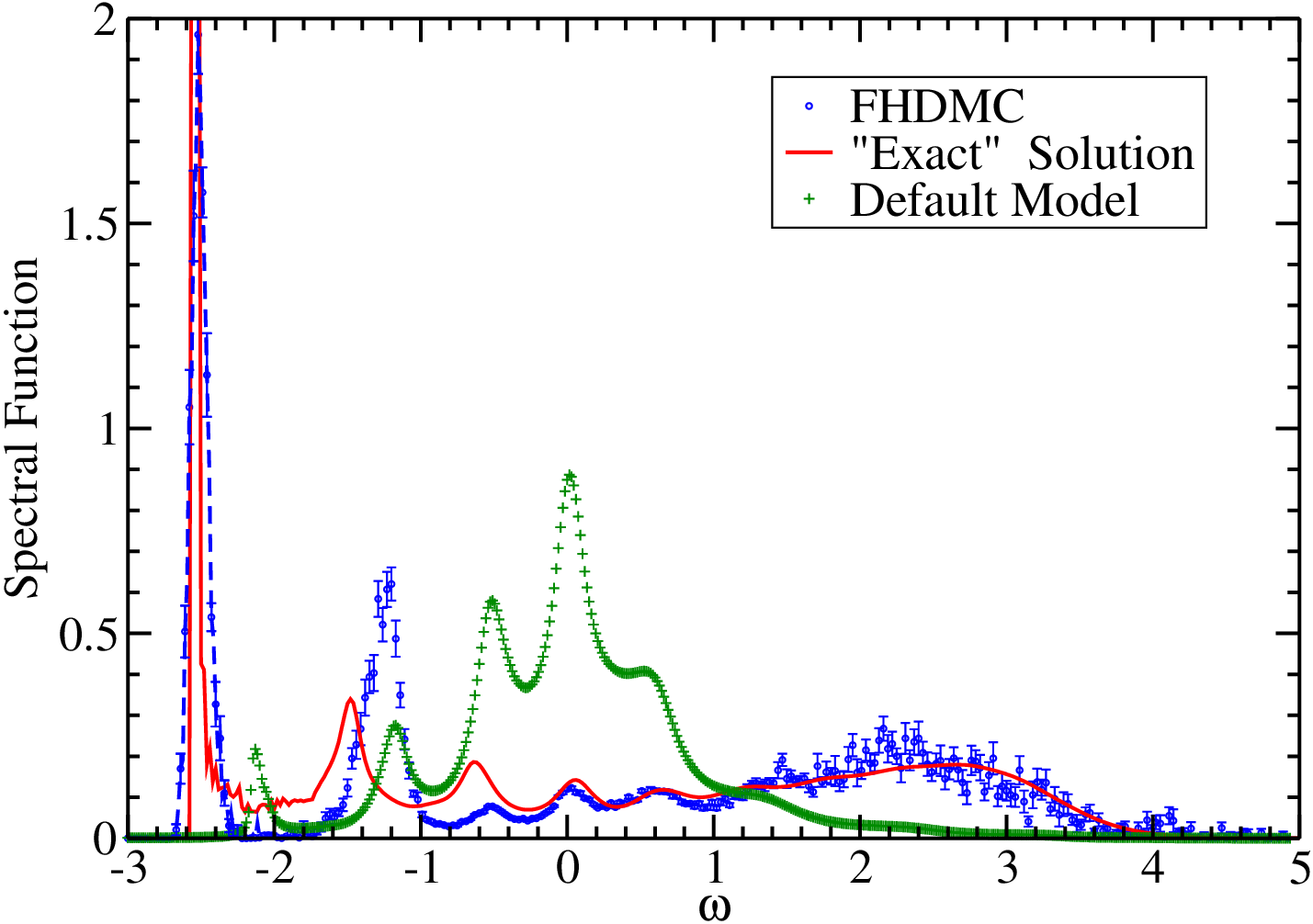}
        }%
    \end{center}
\caption{(Color on-line) Analytic continuation of QMC data 
obtained from the SAI technique using as default model 
the ``exact'' $A(k=0,\omega)$ and $\alpha=1$.
Subfigure (a) is obtained using
the DMC0 data.
Subfigure (b) is obtained using
the standard DMC data.
Subfigure (c) is obtained using
the FHDMC data.
}
\label{fig:4} 
\vskip 0.5 in
\end{figure}

\subsection{Using the ``exact'' as default model}

In Fig.~\ref{fig:2} we present the 
results of the analytic continuation of the data obtained with DMC0,
the standard-DMC, and the FHDMC methods and
the SAI technique, and as default model 
the ``exact'' solution (discussed in the previous section).
Notice that the results using any sets of data 
are very close to the
``exact'' solution and independent of  the ``temperature'' 
parameter $\alpha$ which we used.
In general at higher value of $\alpha$ the role of the default model 
is influencing more the result of the analytic continuation.

\subsection{Using the flat distribution as default model}

Since in practice an exact solution is not available to use as the
default model, we wish to examine the extent to which this method can be used
for the analytic continuation.  
Therefore, we would like to test the method when we use a default model
that is different from the exact answer. Towards this goal, 
we have used the two different default models discussed in the previous Section.
First we use the flat distribution given by Eq.~\ref{flat}.

Fig.~\ref{fig:3} presents the results of the application of the
SAI method using $\alpha=1$. The results for lower values of 
$\alpha$ are very similar. The DMC0 method yields the results 
illustrated in  Fig.~\ref{fig:3}(a)). 
The standard DMC method yields the results presented in 
Fig.~\ref{fig:3}(b)), while 
the FHDMC approach yields the results illustrated in 
Fig.~\ref{fig:3}(c)).
We least-squared fitted the DMC0, the 
standard-DMC, and the FHDMC 
data on  $G(\tau)$  to the form
\begin{eqnarray}
 G(\tau) =  Z_1 e^{-\omega_1 \tau} + Z_2 e^{-\omega_2 \tau} + Z_3 e^{-\omega_3 \tau},
\label{fit}
\end{eqnarray}
and the values of the parameters $Z_{1,2,3}$ and $\omega_{1,2,3}$ found by the 
fit are listed in Table~\ref{table1}.    
The up-triangles in each of the
subfigures show the values of the three frequencies found by the fits.
Notice that the lowest two agree with the lowest two peaks found by the SAI
method. In addition, the third frequency $\omega_3$ approximately accounts for
the broad feature of the 
$A(\omega)$ at high frequencies with approximately the correct spectral weight which is taken into account by the value of $Z_3$. 
The value of $Z_1$, which corresponds to the residue of the lowest 
frequency quasiparticle peak, agrees
rather well with the integral of $A(\omega)$ around the quasiparticle peak;
thus, this agrees with the residue
of the quasiparticle peak found in Ref.~\cite{Liu2}, which  
is approximately 0.2.

\begin{table}[htp]
\begin{tabular}{|p{38pt}|p{20pt}|p{20pt}|p{20pt}|p{25pt}|p{25pt}|p{25pt}|} 
\hline
Method & $Z_1$ & $Z_2$  & $Z_3$ & $\omega_1$      & $\omega_2$  & $\omega_3$  \\ \hline
DMC0 & 0.27 & 0.26  & 0.47 & -2.516      & -1.14  & 2.01  \\ \hline
SDMC & 0.24 & 0.19  & 0.56 & -2.538      & -1.64  &  1.43  \\ \hline
FHDMC & 0.23 & 0.19  & 0.56 & -2.538      & -1.656  & 1.45  \\ \hline
\end{tabular}
\caption{The results of fitting the DMC0, the 
standard-DMC (SDMC),  and the FHDMC data on $G(\tau)$  to the 
formula given by Eq.~\ref{fit}.}
\label{table1}
\end{table}

We notice that the results of the analytic continuation cannot reproduce
the details of the exact solution even by changing the value of $\alpha$
to a significantly lower value. The overall 
performance of the standard DMC and the FHDMC methods are approximately 
the same and
significantly better than the results obtained with the DMC0 method. Notice, 
however, that the width of the lowest energy peak is a little closer to 
the exact when we used the FHDMC data. 

\subsection{Using $A_{(0,0)}(\omega)$ as default model}

The default model used in the previous subsection assumes no a priori 
knowledge about the features of $A_{\vec k}(\omega)$. Therefore, it heavily relies
on minimizing $\chi^2$. In many cases in practice, however, we have some partial information from either approximate analytic or semi-analytic 
(for example, perturbation
expansion) or numerical techniques, such as from the density functional theory 
in electronic structure calculations. It is typical to know the approximate
location of the peaks, however, it is much harder to know
the spectral weight of these peaks. 
Next we will use the ``exact'' spectral function  $A_{(0,0)}(\omega)$  obtained
for $\vec k =(0,0)$ as the default model in the SAI approach to invert
the DMC0, the standard-DMC, and the FHDMC data on $G_{({{\pi} \over 2},{{\pi} \over 2})}(\tau)$.
As can be seen from Fig.~\ref{fig:4}, most of the peaks of the
default model are close to those of the ``exact'' solution. The lowest energy 
peak is in the wrong place and the relative spectral weights (i.e., the
heights of the peaks) are very different than those in the ``exact'' solution.

In Fig.\ref{fig:4}  we present the results of 
applying the SAI method
to data obtained with DMC0, the standard DMC, and the FHDMC method
using $\alpha=1$ and as
default model the above discussed ``incorrect'' solution which 
corresponds to $\vec k=(0,0)$ instead of the correct value of 
$\vec k=(\pi/2,\pi/2)$.
Fig.\ref{fig:4}(a)  presents the results obtained using the DMC0 data.
In Fig.\ref{fig:4}(b) and Fig.~\ref{fig:4}(c)  we present the results of 
applying the SAI method
to data obtained with the standard DMC and the FHDMC method respectively.
The default model is shown by the green open circles.
Notice that the overall performance of the standard DMC and the FHDMC 
methods is approximately the
same and significantly better than that of the DMC0 data. 
Notice, however, that the width of the lowest energy quasiparticle peak
when using the FHDMC data is somewhat smaller and closer to the ``exact'',
as compared to that obtained with the standard DMC data. 
Also  note that the location of the second peak is closer to the
``exact'' location when using the FHDMC data.

These differences in the degree of approximating the low-energy features
of the spectral function can be explained when we compare the
relative error in $G(\tau)$ as obtained from the two methods as 
was done in Fig.~\ref{fig:1}(b). 
Fig.~\ref{fig:1}(b) compares our results for the relative error 
for the $G_{\vec k}(\tau)$ as a function
of  $\tau$ as obtained with the DMC0, FHDMC and standard-DMC
for the approximately the same amounts of CPU time. 
For the case of DMC0 we have chosen $\tau_{max}=3.8$ in order to 
obtain a comparable error in $G(\tau)$ with the other two methods.
First notice that for a fixed value of $\tau_{max}$ and for a fixed 
number of Monte Carlo steps in the DMC0 method, 
the relative error is weakly dependent of $\tau$
yielding an average value $\sigma_r(\tau_{max})$ for the relative error.
However, for fixed number of Monte Carlo steps, this average relative error  
$\sigma_r(\tau_{max})$ in the DMC0 method grows exponentially with 
$\tau_{max}$. By choosing $\tau_{\max}=3.8$ we have optimized the quality of the
analytic continuation. 
While the greater range of $\tau$  allows us to obtain better
resolution for small values of $\omega$, if we increase the value of 
$\tau_{\max}$ beyond the above chosen value, 
the errors grow to the point that the quality of the
analytic continuation becomes much worse. 
This is the reason for the superiority of the
standard DMC and FHDMC methods. 

We further note that
the error in the standard-DMC method at short time $\tau$ is 
much better than the error in the FHDMC data. However, the
situation reverses at long imaginary time. This helps the extraction
of the low-energy physics of the problem. In the case of the
present example, the crossing point of the relative error between the
two methods occurs at $\tau \sim 8$, which implies that the features
of the spectral function obtained with the two methods will differ
for $\delta \omega/t < 1/8$, (where $\delta \omega$ is measured from the
lowest energy state).  For the present problem this seems like a narrow range,
however, in general, for other problems where we are interested in the
low energy excitations it can be more important.

\section{Conclusions}
\label{conclusion}
We have shown that by combining flat 
histogram techniques\cite{Berg,Wang-Landau1,Wang-Landau2,Oliviera}
and the QMC method in such a way to make the histogram of an imaginary-time
correlation function $G(\tau)$ flat, we are able
to carry out accurate analytic continuation to real time with better 
degree of accuracy as compared to any quantum Monte Carlo method 
which does not use the flat histogram idea.

This has been demonstrated using the DMC method which can be
modified to simulate an exactly soluble problem. This is within the
flexibility of the DMC method, because we can select the set of
Feynman diagrams to sample.  Thus, by selecting the diagrammatic series to
be the one which takes into account all the diagrams which contribute to
the single-hole spectral function of the 2D $t-J$ model within the NCA approximation,
we can compare the results of the analytic continuation directly 
to the ``exact'' solution\cite{Liu1,Liu2}.
To carry out the analytic continuation 
to real time, we have used the SAI\cite{SAI} approach, a generalization
of the maximum entropy method. 

From this paper we draw the following main conclusions:

I) The main point of the present paper is the following.
We find that if the QMC data are obtained by a technique in which 
the histogram of $G(\tau)$ is more or less flat, such as the FHDMC or  
the standard DMC,
the results of the analytic continuation improve significantly as compared
to the results obtained by using a QMC method (such as the DMC0 method) in 
which the flat histogram idea has not been applied.
That is because we can approach large imaginary time with significantly 
small statistical error.

II) This paper provides benchmarking on the power of the FHDMC and standard DMC
methods to provide data for analytic continuation to real time using
an exactly soluble model. In particular comparing these two methods, 
we find that:

\begin{itemize}

\item The standard DMC method requires a value of $\mu$ in order to approximate
the long-time behavior of $G(\tau)$ and this can be used to effectively make
the histogram of $G(\tau)$ approximately flat. 
Thus, the standard DMC method requires some amount of prior work in 
order to determine the optimum value of $\mu$.
The FHDMC method makes the histogram of $G(\tau)$ flat automatically 
without requiring any such a priori knowledge or prior work. 

\item The error on the short-time behavior of $G(\tau)$ is significantly
smaller in the case of standard DMC. 

\item The error on the long-time behavior of $G(\tau)$ is 
smaller in the case of FHDMC. This allows us to obtain somewhat better quality
data on the spectrum and spectral weight of the very low-energy excitations.
This aspect of the FHDMC method seems to give some advantage 
over the standard DMC, when we are interested in the very low-energy 
excitations of a given system. 

\end{itemize}
 
III) Lastly, we exploit the fact that we know the ``exact'' solution to this 
problem to  draw the following additional conclusions from the present
calculations:

\begin{itemize}

\item When we use as default model, a distribution which is close to
the exact, the result of the analytic continuation is very close to the
exact result.

\item If the correct answer is completely unknown and we use as
default model the flat distribution, the result for the location for 
and the residue of the lowest energy quasiparticle
peak is very close to the exact value, and the location and width of the
second lowest peak is also in agreement with the ``exact'' answer.
The details of the spectral function cannot be correctly captured by
the analytic continuation using a flat default model.

\item If we approximately know some aspects of the solution, i.e., 
we use a default
model which shares some features of the correct solution 
(such as the location of some of the
peaks, but not their relative strength in the spectral function) 
the analytic continuation tends to correct the strength of the peaks
and tends to move the incorrectly located peaks of the default model
towards the correct positions.

\end{itemize}

The main point of the present paper is that, in general, 
if we incorporate the flat histogram idea with any QMC technique, in such 
a way to 
make the histogram of $G(\tau)$ flat, the results of the analytic 
continuation to real time
should be of better statistical quality.

\vskip 0.2 in
\section{Acknowledgments}
This work was supported in part by the U.S. National High Magnetic Field
Laboratory, which is partially funded by the NSF DMR-1157490 and 
the State of Florida.

\end{document}